\documentclass[journal]{IEEEtran}
\usepackage{amsmath}
\usepackage{amsfonts}
\usepackage{amssymb}
\usepackage{fixltx2e}
\usepackage{xcolor}
\definecolor{Maroon}{cmyk}{0, 0.87, 0.88, 0.32}
\usepackage{cite}
\usepackage{graphicx}
\usepackage{enumitem}
\usepackage{mathrsfs}
\usepackage{comment}
\usepackage[font=scriptsize]{caption}
\usepackage{subcaption}
\usepackage{overpic}
\usepackage{stackengine}
\usepackage{tabularx}
\usepackage{pict2e}

\usepackage{mathtools}

\DeclarePairedDelimiter\floor{\lfloor}{\rfloor}

\def\diag{\mathrm{diag}}

\def\sinc{\mathrm{sinc}}
\def\Htran{\mbox{\tiny $\mathrm{H}$}}
\def\Ttran{\mbox{\tiny $\mathrm{T}$}}
\def\imagunit{\mathsf{j}} 

\newcommand{\vers}[1]{{\bf \hat{#1}}}

\newcommand{\vect}[1]{{\bf{#1}}}

\newcolumntype{K}{>{\centering\arraybackslash$}X<{$}}

\begin{document}

\title{\huge Performance analysis of WDM in LoS communications with arbitrary orientation and position}

\author{Antonio A. D'Amico, Luca Sanguinetti,~\IEEEmembership{Senior~Member,~IEEE}, Merouane Debbah,~\IEEEmembership{Fellow,~IEEE}\vspace{-0.7cm}

\thanks{A. A. D'Amico and L. Sanguinetti are with the Dipartimento di Ingegneria dell'Informazione, University of Pisa, 56122 Pisa, Italy. M. Debbah is with the Mathematical and Algorithmic Sciences Lab, Huawei Technologies Co., Ltd., 92100 Boulogne Billancourt, France. 
}}

\maketitle

\begin{abstract}
This letter considers the wavenumber-division-multiplexing (WDM) scheme that was recently presented in~\cite{sanguinetti2022WDM} for line-of-sight communications between parallel (i.e., side-by-side) spatially-continuous electromagnetic segments. Our aim is to analyze the performance of WDM, combined with different digital processing architectures, when the electromagnetic segments have an arbitrary orientation and position. To this end, we first show how the general electromagnetic MIMO (multiple-input multiple-output) model from~\cite{sanguinetti2022WDM} can be particularized to the case of interest and then use numerical results to evaluate the impact of geometric parameters (e.g., horizontal and vertical distances, azimuth and elevation orientations). It turns out that WDM performs satisfactorily also when the transmit and receive segments are not in boresight direction to each other.
\end{abstract}

\begin{IEEEkeywords}
	Spatially-continuous electromagnetic segments, holographic communications, wavenumber-division-multiplexing, line-of-sight communications.
\end{IEEEkeywords}

\vspace{-0.4cm}
\section{Introduction}
Classical MIMO (multiple-input multiple-output) literature teaches us that spatial multiplexing is inevitably compromised in environments with not-rich scattering~\cite{TseBook}. This is true under the assumption that the wavefronts of radiated waves can be approximated as locally planar over the entire antenna arrays. This assumption breaks down when large arrays combined with high carrier frequencies are used. Since the wavelength reduces dramatically and the transmission range tends to be short, the wave curvature over the array is no longer negligible. This opens the door for spatial multiplexing even in situations with little or no scattering~\cite{Bohagen-2009,Madhow-2011,Heedong-2021}. 

The potential benefits of large arrays combined with high carrier frequencies is driving a flurry of research activity at the intersection of information theory and electromagnetics, with the promise of concepts such as holographic MIMO communications~\cite{Sanguinetti2021}, known also as continuous-aperture MIMO (CAP-MIMO) communications~\cite{sayeed2013a}, or large intelligent surface (LIS) communications~\cite{Edfors2018}. The dream of these concepts is to achieve full control of the electromagnetic field that is generated and sensed in a communication system by means of spatially-continuous electromagnetic surfaces. From the technological point of view, metamaterials represent appealing candidates toward the creation of such surfaces~\cite{Tretyakov2015}. 

Recently, in~\cite{sanguinetti2022WDM} we considered two parallel (i.e., side-by-side) electromagnetic segments and directly generated the spatially-continuous transmit currents and received fields through a series of Fourier basis functions. This leads to a wavenumber-division multiplexing (WDM) scheme, which modulates the transmitted symbols onto orthogonal spatial beams, e.g.,~\cite{sayeed2013a}. The scheme is not optimal but can be efficiently implemented. Due to the non-finite support of the electromagnetic channel, WDM cannot provide non-interfering communication modes. Different digital processing architectures are thus used to deal with the interference. The analysis in~\cite{sanguinetti2022WDM} is exclusively carried out under the assumption that the electromagnetic segments are side-by-side. However, this may not be the case in practical scenarios.

\begin{figure}[t!]\vspace{-0.4cm}
	\centering 
	\begin{overpic}[width=0.7\columnwidth,tics=10]{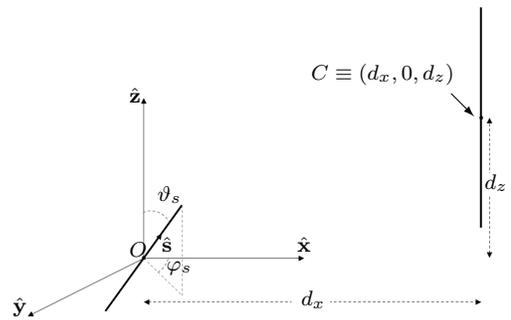}
	\put(61,51){\footnotesize $C \equiv (d_x, 0, d_z)$}
	\put(91,48){\vector(1, -1){5}}
	\put(28,25){\footnotesize ${\vartheta_s}$}
		\put(30,10){\footnotesize ${\varphi_s}$}
		\put(29,14){\footnotesize $\hat{\bf s}$}
	\put(58,14){\footnotesize $\hat{\bf x}$}
	\put(58.9,2.5){\footnotesize $d_x$}
	\put(98.2,27.6){\footnotesize $d_z$}
	\put(-3,1){\footnotesize $\hat{\bf y}$}
	\put(22,46){\footnotesize $\hat{\bf z}$}
	
	\put(22,13){\footnotesize $O$}
\end{overpic} 
	\caption{Geometry of the communication system. The transmit and receive line segments have an arbitrary arrangement.}\vspace{-0.4cm}
	\label{figure_linear_arrays} 
\end{figure}

The aim of this letter is to extend~\cite{sanguinetti2022WDM} in two directions. Firstly, we provide the WDM signal model for arbitrary arrangements of electromagnetic segments; see Fig.~\ref{figure_linear_arrays}. Secondly, we evaluate by means of analytical and numerical results how different arrangements impact the radiation pattern and the spectral efficiency of various communication architectures. In particular, we assess the effects of the geometric parameters in Fig.~\ref{figure_linear_arrays}, such as the horizontal distance $d_x$, the vertical misalignments $d_z$, the angles $\vartheta_s$ and $\varphi_s$, on the received power and on the interference pattern between the communication modes. Our analysis reveals that WDM performs satisfactorily as the above parameters vary within practical ranges.

\begin{figure*}\vspace{-0.4cm}
\begin{equation}\tag{13}
\label{matRot}
\mathbf{Q}=\left[\begin{array}{ccc} \sin^2 \varphi_s+\cos^2 \varphi_s\cos \vartheta_s & -\sin \varphi_s \cos  \varphi_s(1-\cos \vartheta_s) & -\cos \varphi_s \sin \vartheta_s \\ -\sin \varphi_s \cos  \varphi_s(1-\cos \vartheta_s) & \cos^2 \varphi_s+\sin^2 \varphi_s\cos \vartheta_s & -\sin \varphi_s \sin \vartheta_s \\ \cos \varphi_s \sin \vartheta_s & \sin \varphi_s \sin \vartheta_s & \cos \vartheta_s \end{array}\right]
\end{equation}
\hrule\vspace{-0.2cm}
\end{figure*}
\vspace{-0.2cm}
\section{Review of Electromagnetic MIMO Model}
Consider two volumes of arbitrary shape and position that communicate by means of electromagnetic waves through an isotropic, homogeneous, unbounded medium. We consider only monochromatic sources\footnote{The sources are bounded, which guarantees a bounded radiated power.} and electric fields. {An electric current density ${\bf j}({\bf s})$ at any arbitrary source point ${\bf s}$, inside the source volume $V_s$, generates an electric field ${\bf e}(\vect{r})$ in [V/m] at a generic location $\vect{r}$ of the receiving volume $V_r$.}
\vspace{-0.3cm}
\subsection{The received electric field}
The electric field ${\bf y}({\bf r})$ observed in the receiving volume $V_r$ is the sum of the information-carrying electric field ${\bf e}(\vect{r})$, produced by ${\bf j}({\bf s})$, and a random noise field ${\bf n}(\vect{r})$, i.e.,
\begin{align} \label{eq:IO}
{\bf y}(\vect{r})  = {\bf e}(\vect{r}) + {\bf n}(\vect{r})
\end{align}
where \cite[Eq.~(1.3.53)]{ChewBook}
\begin{equation}
\label{FieldCurrent}
\begin{split}
{\bf e}(\vect{r})&=\imagunit \kappa Z_0 \int_{V_s} {\bf g}(\vect{r},\vect{s}) {\bf j}(\vect{s}) \, {\rm d}\vect{s}\\
\end{split}
\end{equation}
$\kappa=\omega/c=2\pi/\lambda$ is the wavenumber ({with $c$ and $\lambda$ being the speed of light and the wavelength, respectively}), $Z_0 = 376.73$\,[Ohm] is the free-space intrinsic impedance,
${\bf g}(\vect{r},\vect{s})$ is the dyadic Green's function~\cite[Eq.~(1.3.51)]{ChewBook}, and ${\bf n}(\vect{r})$ accounts for the electromagnetic interference (EMI) generated by external sources~\cite{Wallace2008}. 
In the evaluation of fields away from the sources, where $ \|\vect{r}-\vect{s}\| \gg \lambda$, ${\bf g}(\vect{r},\vect{s})$ can be approximated\footnote{The \textit{far-field approximation} in electromagnetic propagation corresponds to the radiated field that falls off inversely as the distance apart $\|\vect{r}-\vect{s}\|$. Hence, its power follows the inverse square law.} as~\cite{Poon}
\begin{equation}
\label{DyadicGF.2}
\begin{split}
\vect g(\mathbf{r},\mathbf{s}) \approx \dfrac{1}{4 \pi} \dfrac{e^{ \imagunit \kappa \|\vect{r}-\vect{s}\|}}{\|\vect{r}-\vect{s}\|}  \left(\mathbf{I}_3-\widehat{\bf p} \widehat{\bf p}^{\Htran}\right)\end{split}
\end{equation}
where $\widehat{\bf p} = {\bf p}/||{\bf p}||$, $\bf{p}=\bf{r}-\bf{s}$ and {$\mathbf{I}_3$ is the identity matrix.}

\vspace{-0.3cm}
\subsection{The MIMO model}
We assume that the current sources are expanded using a set of orthonormal vector functions $\{{\boldsymbol\phi}_m(\vect{s}); m=1,\ldots,N\}$ such that ${\bf j}(\vect{s}) = \sum_{m=1}^N \xi_m {\boldsymbol \phi}_m(\vect{s})$.
Similarly, the field ${\bf y}(\vect{r})$ in~\eqref{eq:IO} is projected onto an output space spanned by a set of orthonormal vector functions $\{{\boldsymbol \psi}_n(\vect{r}); n=1,\ldots,N\}$. Hence, the spatial samples $\{y_n; n =1,\ldots,N\}$ are given by~\cite[Eq.~(24)]{sanguinetti2022WDM} 
\begin{align}
y_n = \sum_{m=1}^N H_{nm} x_m + z_n\label{eq:IO-discrete}
\end{align}
where $x_m = \imagunit \kappa Z_0 \xi_m$ are the effective input samples, $z_n =  z_n^{({\rm emi})} + z_n^{({\rm hdw})}$ with $z_n^{({\rm hdw})}\sim \mathcal{NC}(0,\sigma^2_{{\rm hdw}})$ being the noise of hardware nature and $z_n^{({\rm emi})} = \int_{V_r} {\boldsymbol \psi}_n^{\Htran}(\vect{r}) {\bf n}(\vect{r})  d\vect{r}$
while
\begin{align}
H_{nm} =\int_{V_r}\int_{V_s} {\boldsymbol \psi}_n^{\Htran}(\vect{r}) {\bf g}({\bf r}, {\bf s}) {\boldsymbol \phi}_m(\vect{s}) d \vect{r} d \vect{s}\label{eq:H_ni}
\end{align}
represents the \textit{coupling coefficient} between the source mode $m$ and the reception mode $n$~\cite{Miller:00}. Letting ${\bf y} = [y_1,\ldots,y_N]^{\Ttran}$ and ${\bf x} = [x_1,\ldots,x_N]^{\Ttran}$, we may rewrite~\eqref{eq:IO-discrete} in matrix form 
\begin{align}\label{eq:MIMO_channel_model}
{\bf y} = {\bf H}{\bf x} + {{\bf z}}
\end{align}
where ${\bf H}\in \mathbb{C}^{N\times N}$ is the channel matrix and 
${{\bf z}}=[z_1,\ldots,z_N]^{\Ttran}\sim \mathcal{N}_{\mathbb{C}} ({\bf 0}_N, {\bf C})$, with ${\bf C} = \sigma^2_{{\rm emi}}{\bf R} + {\sigma^2_{{\rm hdw}}}{\bf I}_N$,
\begin{align}\label{eq:correlation_matrix}
\left[ {\bf R}\right]_{nm}= \iint_{V_r}  \rho(\vect{r}-\vect{r}^\prime) {\boldsymbol \psi}_n^{\Htran}(\vect{r}){\boldsymbol \psi}_m(\vect{r}^\prime) d\vect{r}d\vect{r}^\prime
\end{align}
and $\rho(\vect{r})$ is the EMI spatial correlation function that depends on the power angular density~\cite{sanguinetti2022WDM}.
With isotropic propagation conditions, we have that $\rho(\vect{r})=\sinc \left({2||{\bf r}||}/{\lambda}\right)$~\cite[Eq.~(20)]{sanguinetti2022WDM}.

\section{WDM Model with Arbitrary Orientation}
The MIMO model~\eqref{eq:MIMO_channel_model} is valid for volumes of arbitrary shape and position, and any arbitrary set of functions $\{{\boldsymbol\phi}_m(\vect{s}); m=1,\ldots,N\}$ and $\{{\boldsymbol \psi}_n(\vect{r}); n=1,\ldots,N\}$. Although possibile, their optimal design leads to a communication system of prohibitive complexity~\cite{Miller:00}. {Next,~\eqref{eq:IO-discrete} will be specialized to the system depicted in Fig.~\ref{figure_linear_arrays} and the WDM scheme presented in~\cite{sanguinetti2022WDM}, which makes use of Fourier basis functions. Although suboptimal, this choice makes the communication system behave as OFDM (orthogonal-frequency division-multiplexing) in time-domain dispersive channel, and also provides it a clear physical interpretation. In fact, the use of Fourier basis functions in the wavenumber-domain produces orthogonal spatial beams towards specific angular directions, e.g.,~\cite{sayeed2013a}.}
\subsection{System model}
In Fig.~\ref{figure_linear_arrays}, {the transmitting source is a linear segment of length $L_s$} that lies along the direction of the unitary vector 
\begin{equation}
\vers s=[\cos \varphi_s \sin \vartheta_s, \sin \varphi_s \sin \vartheta_s, \allowbreak \cos \vartheta_s]^T
\end{equation}
where $\varphi_s$ and $\vartheta_s$ are azimuth and polar angles, respectively.  We assume that the source center is coincident with the origin $O$ of the Cartesian coordinate system $O \vers x \vers y \vers z$. Accordingly, the linear region $V_s$ is the set of points $(s_x,s_y,s_z)$ given by
\begin{align}
(\rho_s \cos \varphi_s \sin \vartheta_s, \rho_s \sin \varphi_s \sin \vartheta_s, \rho_s \cos \vartheta_s)
\end{align}
with $|\rho_s| \le L_s/2$.  Without loss of generality, we assume that the receiving line segment is directed along $\vers z$ and occupies the linear region, given by
\begin{equation}
\label{ReceiveRegion}
V_r=\{(r_x,r_y,r_z)|r_x=d_x, r_y=0, |r_z-d_z| \le L_r/2\}\}
\end{equation} 
where $\sqrt{d_x^2+d_z^2}$ denotes the distance between the centers. {We call $C$ the center of the segment.}
{In the system shown in Fig.~\ref{figure_linear_arrays}, the current density ${\bf j}(\vect{s})$ is directed along $\vers s$. Hence, the vector function ${\boldsymbol\phi}_m(\vect{s})$  is also directed along $\vers s$ and can be written in the form:}
\begin{align}\label{eq:BI}
{\boldsymbol\phi}_m(\vect{s})= \phi_m(s_{z'}) \delta(s_{x'}) \delta(s_{y'}) \vers s
\end{align}
where $\phi_m(s_{z'})$ are scalar functions and 
\begin{equation}
\label{newsystem}
\left(\begin{array}{c} s_{x'} \\ s_{y'}  \\ s_{z'} \end{array}\right)={\mathbf Q} \left(\begin{array}{c} s_x \\ s_y \\ s_z \end{array}\right)
\end{equation}
where $\mathbf{Q}$ is the rotation matrix given in~\eqref{matRot}. Notice that this matrix performs a change in the coordinate system in which the new axis $\vers z'$ coincides with $\vers s$ and the source region becomes $V_s=\{(s_{x'},s_{y'},s_{z'})|s_{x'}= 0, s_{y'}= 0, |s_{z'}|\le L_s/2\}$. 

\begin{figure*}\vspace{-0.5cm}
\begin{align}
\label{gz} \tag{20}
g_z({\bf u})&= g_z(u_x,u_y,u_z)= \dfrac{e^{ \imagunit \kappa \|{\bf u}\|}}{4 \pi \|{\bf u}\|^3} \left[ -u_x u_z \cos \varphi_s \sin \vartheta_s - u_y u_z \sin \varphi_s \sin \vartheta_s  +(u_x^2+u_y^2) \cos \vartheta_s \right]
\end{align}
\hrule\vspace{-0.3cm}
\end{figure*}
\vspace{-0.25cm}
\subsection{Signal model}
{To overcome the prohibitive implementation complexity of the optimal basis functions, e.g.~\cite{Miller:00}, in~\cite{sanguinetti2022WDM} Fourier basis functions are used for representing current sources and electric fields.} Specifically, $\phi_n(s)$ takes the form  
\setcounter{equation}{13}
\begin{equation}
\label{fi_n}
\phi_n(s) =  \begin{cases}
\;\;\frac{1}{\sqrt{L_s}}e^{\imagunit \kappa_n s}, &   |s| \le L_s/2 \\
\;\; 0, & \text{elsewhere}
\end{cases}
\end{equation}
where
\begin{equation}
\label{kappa_m}
\kappa_n=\dfrac{2 \pi}{L_s} \left(n-\dfrac{N+1}{2}\right) 
\end{equation}
for $n=1,\ldots,N$, denotes the \textit{spatial frequency}. As shown in~\cite{sanguinetti2022WDM}, the maximum dimension of the input signal space depends on its wavenumber spectrum and is thus given by
\begin{align}\label{eq:max_number_of_modes}
N\le {N_{\max} = 2\floor*{ \frac{L_s}{\lambda}}+1}.
\end{align}
Following~\cite{sanguinetti2022WDM}, at the receiver we only consider the component of electric field along the receive direction, i.e., $z-$axis. From~\eqref{ReceiveRegion}, it thus follows
\begin{align}\label{eq:B0-rx_1}
{\boldsymbol{\psi}}_n({\bf r}) = \psi_n(r_z) \delta(r_x-d_x) \delta(r_y){\hat {\bf z}}
\end{align}
where $\psi_n(r_z)$ are scalar functions
\begin{align}\label{eq:B0-rx}
\psi_n(r) =  \begin{cases}
\;\;e^{\imagunit \kappa_n r}, &   |r-d_z|\le L_r/2 \\
\;\; 0, & \text{elsewhere}
\end{cases}
\end{align}
for $n=1,\ldots,N$. The elements of $\bf H$ and ${\bf R}$ are then computed by using \eqref{eq:BI} and \eqref{eq:B0-rx}, into \eqref{eq:H_ni} and \eqref{eq:correlation_matrix}. Specifically, the coupling coefficients in \eqref{eq:H_ni} assume the following form
\begin{equation}
\label{Hnm.2}
H_{nm}=\!\!\!\!\!\!\int\limits_{d_z-L_r/2}^{d_z+L_r/2} \int\limits_{-L_s/2}^{L_s/2} \!\!g_z({\bf r}-s_{z'}{\vers s}) \phi_m (s_{z'}) \psi^{\ast}_n (r_z) \, {\rm d} s_{z'} \, {\rm d} r_{z}
\end{equation}
where $g_z({\bf u})$ is given in \eqref{gz}. The entries of ${\bf R}$ in \eqref{eq:correlation_matrix} are
\setcounter{equation}{20} 
\begin{align}\label{eq:correlation_matrix.2}
\left[ {\bf R}\right]_{nm}= \iint\limits_{-L_r/2}^{\hphantom{-}L_r/2}  \rho(r_z^{\prime}-r_z) \psi_n^{\ast}(r_z)\psi_m(r_z^{\prime}) {\rm d} r_z {\rm d} r_z^{\prime}.
\end{align}

\vspace{-0.4cm}
\subsection{Radiation pattern}
\begin{figure}
\begin{center}
\includegraphics[width=0.43\textwidth]{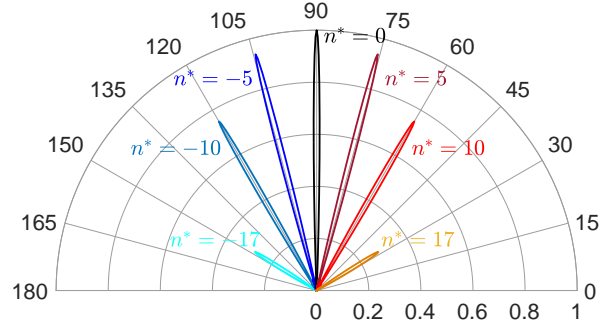}
\caption{Radiation pattern along a vertical plane for different values of $\kappa_n$.}\vspace{-0.7cm}
\label{FigRP0}
\end{center}
\end{figure}

{The use of the exponential functions \eqref{fi_n} produces spatial beams directed towards an angular direction (\emph{relative to the antenna axis}) given by $\cos^{-1} (\kappa_n/\kappa)$, e.g.,~\cite{sayeed2013a}.} To show this, consider the electric field generated by the current distribution 
\begin{equation}\label{J.1}
{\bf j}(\vect{s})= \xi \phi_n(s_{z'}) \delta(s_{x'}) \delta(s_{y'}) \vers s 
\end{equation}
as defined by \eqref{eq:BI} and \eqref{newsystem}, with $\phi_n(s_{z'})$ given by~\eqref{fi_n}. Plugging~\eqref{J.1} into \eqref{FieldCurrent}, from~\eqref{DyadicGF.2} we obtain
\begin{equation}
\label{Efield.1}
{\bf e}({\bf r})=\dfrac{\imagunit \kappa Z_0 \xi}{4 \pi} \int\limits_{-L_s/2}^{L_s/2} \phi_n (s) \dfrac{e^{ \imagunit \kappa \|\vect{p}\|}}{\|\vect{p}\|}  \left(\mathbf{I}-\widehat{\bf p} \widehat{\bf p}^{\Htran}\right){\vers s}  \, {\rm d} s
\end{equation}
with $\vect{p}={\bf r}-s{\vers s}$. In the limit, as $\| {\bf r}\| \to \infty$, \eqref{Efield.1} can be approximated as (e.g.,~\cite[App. B]{sanguinetti2022WDM})
\begin{equation}
\label{Efield.2}
{\bf e}({\bf r}) \approx  A\left(\|\vect{r}\|\right) {\rm sinc}\left(\dfrac{2L_s}{\lambda}\left(\gamma_n -\cos \bar \theta \right) \right)(\vers s - \cos \bar \theta \vers r)
\end{equation}
where $\cos \bar \theta=\vers r \cdot \vers s$ (i.e., $\bar \theta$ is the angle between $\vers r$ and $\vers s$), $\gamma_n=\kappa_n/\kappa$ and  
\begin{equation}
\label{A}
A\left(\|\vect{r}\|\right)=\dfrac{\imagunit \kappa Z_0 \xi \sqrt{L_s} e^{ \imagunit \kappa \|\vect{r}\|}}{4 \pi \|\vect{r}\|}.
\end{equation}
The normalized radiation pattern is defined as $\mathscr{P}_n(\bar \theta)=\|{\bf e}({\bf r})\|^2/|A\left(\|\vect{r}\|\right)|^2$. Hence, from \eqref{Efield.2} we obtain
\begin{equation}
\label{RadPat}
\mathscr{P}_n(\bar \theta)=(\sin\bar \theta)^2 {\rm sinc}^2\left(\dfrac{2L_s}{\lambda}\left(\gamma_n -\cos \bar \theta \right) \right).
\end{equation}
Assume that $\lambda=0.01$ m and $L_s=0.2$ m such that $N_{\max}=41$. Fig.~\ref{FigRP0} shows a polar plot of $\mathscr{P}_n(\bar \theta)$ for $n=(N+1)/2+n^{\ast}$ when $N=N_{\max}=41$ and $n^{\ast}=0, \pm 5, \pm 10, \pm 17$. The plot shows that the maximum of $\mathscr{P}_n(\bar \theta)$ is attained at $\bar \theta_n = \cos^{-1} (\gamma_n)$, as it follows from \eqref{RadPat}. Its value is equal to $1-\gamma^2_n$, and thus it depends on the spatial frequency $\kappa_n$.

\vspace{-0.3cm}
\subsection{Behaviour of the received electric field}

\begin{figure}\vspace{-0.4cm}
\centering
\begin{subfigure}{.5\textwidth}
\centering
\begin{overpic}[width=0.9\textwidth,tics=1]{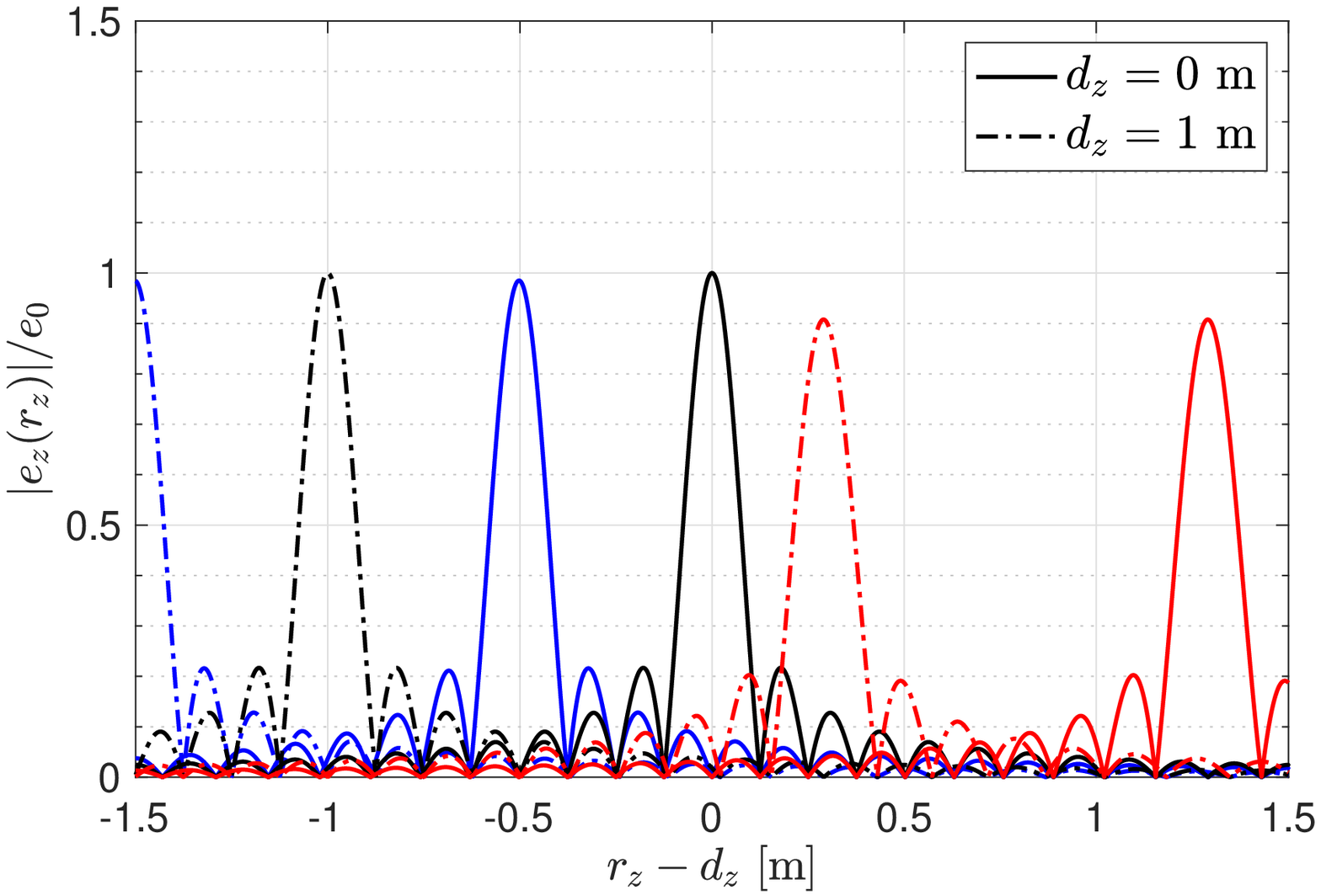} 
\linethickness{0.01 ex}

\put(15,49){\textcolor{blue}{\footnotesize $n=19$}}
\put(20,48){\line(-707, -677){6.8}}
\put(20,48){\line(945, -327){19.0}}

\put(40,49){{\footnotesize $n=21$}}
\put(45,48){\line(-954, -301){19.4}}
\put(45,48){\line(750, -662){7}}

\put(66,43){\textcolor{red}{\footnotesize $n=26$}}
\put(71,42){\line(-962, -271){11.8}}
\put(71,42){\line(974, -228){14.}}
\end{overpic}

\caption{ $\vartheta_s=\varphi_s=0^{\circ}$.}
\label{FigRP1}
\end{subfigure}

\begin{subfigure}{.5\textwidth}
  \centering
\begin{overpic}[width=0.9\textwidth,tics=1]{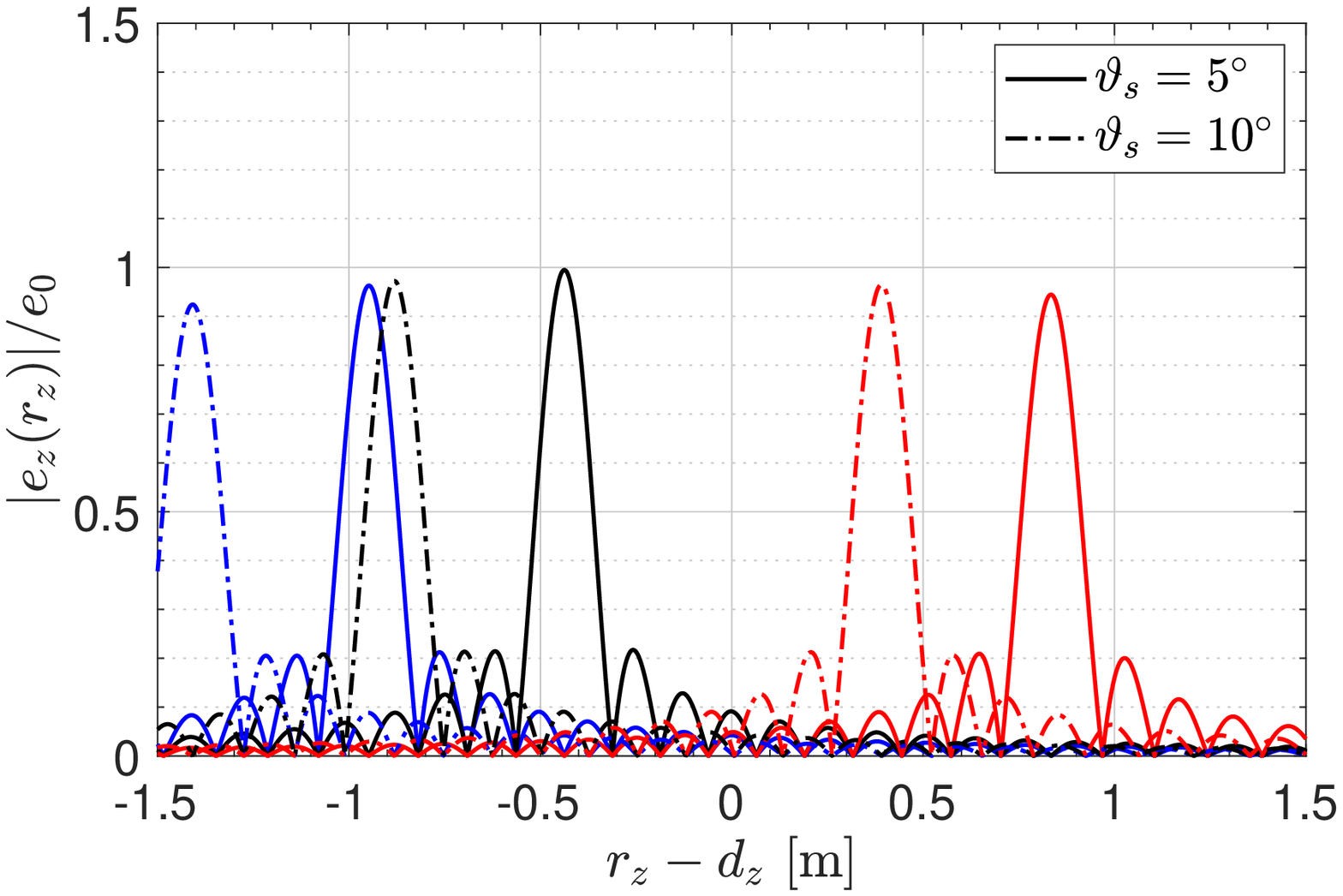} 
\linethickness{0.01 ex}

\put(15,49){\textcolor{blue}{\footnotesize $n=19$}}
\put(20,48){\line(107, -100){7.3}}
\put(20,48){\line(-230, -400){4.6}}

\put(32,49){{\footnotesize $n=21$}}
\put(37,48){\line(-79.26, -62.97){8.1}}
\put(37,48){\line(550, -900){3.8}}

\put(63,43){\textcolor{red}{\footnotesize $n=26$}}

\end{overpic}
\caption{$d_z=0$\,m and $\varphi_s=0^{\circ}$.}
\label{FigRP2}
\end{subfigure}
\caption{Normalized vertical received field for $\kappa_n=19,21,26$.}\vspace{-0.5cm}
\end{figure}

Based on the radiation pattern analysis above, we expect the received field to be strongly affected by the relative position between the source and receive line segments in~Fig.~\ref{figure_linear_arrays}.
To show this, we consider the $z$-component $e^{(n)}_z(r_z)$ of the electric field ${\bf e}({\bf r})$ induced by $\phi_n (s)$. Fig.~\ref{FigRP1} shows $|e^{(n)}_z(r_z)|$ as function of $r_z-d_z$ for $d_z=0$\,m and $d_z=1$\,m when the source segment is vertically oriented, i.e., $\vartheta_s=\varphi_s=0^{\circ}$. We assume $\lambda=0.01$~m,  $L_s=0.2$~m and $L_r=3$~m. Three spatial frequencies are considered, namely $\kappa_{19}=-4 \pi/L_s$ (blue curves), $\kappa_{21}=0$ (black curves) and $\kappa_{26}=10 \pi/L_s$ (red curves). All the curves are normalized to $e_0$ corresponding to the maximum of $|e^{(21)}_z(r_z)|$, obtained for $r_z=0$ (solid black curve). We see that the position of the maximum of $|e^{(n)}_z(r_z)|$, relative to the center of the receive segment, shifts leftward as $d_z $ increases. This can be explained by recalling that each spatial frequency is associated to an angular direction (relative to the antenna axis) given by $\cos^{-1} (\gamma_n)$, as shown in Fig~\ref{FigRP0}. Accordingly, $|e^{(n)}_z(r_z)|$ achieves its maximum in 
\begin{align}
\label{rn}
r^{(n)}_z=\frac{d_x\gamma_n}{\sqrt{1-\gamma_n^2}}
\end{align}
and the corresponding normalized value is approximately given by $(1-\gamma_n^2)^{3/2}$. The distance of the maximum from the center of the receive segment is exactly $r^{(n)}_z-d_z$. Notice that $r^{(n)}_z$ belongs to the receive segment only if $|r^{(n)}_z-d_z| < L_r/2$.

In the general case with $\vartheta_s \ne 0^{\circ}$, $\varphi_s \ne 0^{\circ}$, the directions of maximum radiation (which lie on a cone of vertex $O$ and aperture $2 \bar \theta_n$) intersect the line $(d_x,0,r_z)$ only if $\Delta \ge 0$ with 
\begin{equation}
\label{ }
\Delta=1-\sin^2\varphi_s \sin^2\vartheta_s-\gamma^2_n.
\end{equation}
In this case, one or two intersection points may exist:
\begin{equation}
\label{rn.gen}
r^{(n)}_z=d_x \dfrac{-\cos\varphi_s \sin\vartheta_s \cos\vartheta_s \pm |\gamma_n|\sqrt{\Delta}}{\cos^2\vartheta_s-\gamma^2_n}
\end{equation}
provided that $d_x \cos\varphi_s \sin\vartheta_s + \cos\vartheta_s r^{(n)}_z \ge 0$. In general, the values of $e^{(n)}_z(r_z)$ in the two intersection points are different, as they depend on the distance $\sqrt{d_x^2+(r^{(n)}_z)^2}$. As before, $r^{(n)}_z$ belongs to the receive segment only if $|r^{(n)}_z-d_z| < L_r/2$. Fig.~\ref{FigRP2} shows $|e_z(r_z)|/e_0$ as function of $r_z-d_z$ for $\vartheta_s=5^{\circ}$ and $\vartheta_s=10^{\circ}$, with $\varphi_s=0^{\circ}$, $d_x=5$ m and $d_z=0$\,m. The spatial frequencies are the same as in Fig.~\ref{FigRP1}. We see that when $\vartheta_s$ increases the maximum value of $|e_z(r_z)|/e_0$ shifts leftward. It is attained for $r^{(n)}_z$ as given in \eqref{rn.gen}. If, for example, $n=21$, then $r^{(21)}_z = -d_x \tan \vartheta_s$ and $|e_z(r^{(21)}_z)|/e_0 \approx \cos^2(\vartheta_s)$.

\section{Simulation results}
Numerical results are now used to assess the performance of WDM with different geometric parameters. We assume $\lambda = 0.01$\,m, $L_s = 0.2$\,m and $L_r = 3$\,m. The maximum number of communication modes is ${N_{\max} = 2\floor*{L_s/\lambda}+1}=41$,  with spacing in the wavenumber domain of $2\pi/L_s = 31.41$\,rad/m~\cite{sanguinetti2022WDM}. We assume that all communication modes can be used for transmission, and hence we set $N = N_{\rm {\max}}$.

{The analysis is performed by considering the same digital processing architectures presented in~\cite{sanguinetti2022WDM} and listed in the first column of Table~\ref{tab:scheme}.\footnote{{We refer the interested reader to~\cite{sanguinetti2022WDM} for further details.}}} All the schemes operate through a pre-processing matrix $\bf A$ and a post-processing matrix $\bf B$. Specifically, the transmitted vector is $\bf x = A x'$, with ${\bf x'}=[x'_1,\ldots,x'_N]^{\Ttran}$ and $p_n = \mathbb{E}\{|x'_n|^2\}$. At the receiver, the output $\bf y$ is processed by $\bf B$ to obtain $\bf y'=B^{\Htran} y$. In particular, ${\bf B}={\bf L}^{-\Htran}\widetilde{\bf B}$, where ${\bf L}{\bf L}^{\Htran} = {\bf C}$, and $\widetilde{\bf B}$ is a design matrix that depends on the adopted scheme from Table~\ref{tab:scheme}. Accordingly, we have that ${\bf y'}=\widetilde{\bf B}^{\Htran} \widetilde {\bf H}{\bf A x'} +\widetilde{\bf B}^{\Htran}{\bf z'}$ where $\widetilde {\bf H}= {\bf L}^{-1} {\bf H}$ and ${\bf z'} $ has independent and identically distributed Gaussian entries with ${z}'_n \sim \mathcal N(0,1)$.
The power coefficients $\{p_n;1,\ldots,N\}$ are determined through the water-filling algorithm. This yields $p_n=\max\{ 0,\mu - {1}/{\chi_n}\}$, where $\mu$ is chosen so as to satisfy the power constraint $\sum_{n=1}^N p_n=P$ and $\chi_n$ depends on the particular scheme. The parameters characterizing each scheme are listed in Table~\ref{tab:scheme}, where $\bf U$,  $\bf \Sigma$ and $\bf V$ are the matrices involved in the singular-value decomposition (SVD) of $\widetilde {\bf H} = [\widetilde {\bf h}_1,\ldots, \widetilde {\bf h}_N]$, i.e., $\widetilde {\bf H}=\bf U \Sigma V ^{\Htran}$, and ${\bf P}=\diag(p_1,\ldots,p_N)$. {Notice that the different schemes have different computational complexities, e.g.,~\cite{TseBook}.}

\begin{table}[t]\vspace{-0.4cm}
\centering
\caption{{Parameters of the digital processing architectures from~\cite{sanguinetti2022WDM}}.}
\begin{tabular}{>{\raggedright}m{60 pt}|c|c|c}
Scheme & $\bf A$ & $\bf \widetilde{\bf B}$ & $\chi_n$\\
\hline
\rule{0pt}{12pt}{WDM w/ SVD}  & $\bf V$ & $\bf U$ & $[{\bf \Sigma}]_{nn}^2$ \\
\hline
\rule{0pt}{12pt}WDM w/ MMSE  & ${\bf I}_N$ & $\left( {\bf P} \widetilde {\bf H} \widetilde {\bf H}^{\Htran}+ {\bf I}_N \right)^{-1} \widetilde {\bf H}$ & $\big\|\widetilde{\bf h}_n\big\|^2$\\
\hline
\rule{0pt}{12pt}WDM w/ MR  & ${\bf I}_N$ & $\widetilde {\bf H}$ & $\big\|\widetilde{\bf h}_n\big\|^2$\\
\hline
\rule{0pt}{12pt}WDM  & ${\bf I}_N$ & $-$ & $|[\widetilde{{\bf H}}]_{nn}|^2$\\
\hline
\end{tabular}\vspace{-0.5cm}
\label{tab:scheme}
\end{table}

The different schemes are compared in terms of spectral efficiency (SE). Specifically, the SE of WDM with SVD is
\begin{equation}
\label{SE_SVD}
\mathsf{SE}^{(\rm SVD)}=\sum\limits_{n=1}^{N} \log_2 \left(1+p_n [{\bf \Sigma}]_{nn}^2\right)
\end{equation}
whilst, for the other three schemes, it is computed as $\mathsf{SE}=\sum\nolimits_{n=1}^{N} \log_2 \left(1+\mathsf{SINR}_n\right)$
where
\begin{equation}
\label{SINR}
\mathsf{SINR}_n = \frac{|\widetilde{\bf b}_n^{\Htran}\widetilde{\bf h}_n|^2 p_n}{\sum\limits_{m=1,m\ne n}^{N}|\widetilde{\bf b}_n^{\Htran}\widetilde{\bf h}_m|^2 p_m+ \widetilde{\bf b}_n^{\Htran}\widetilde{\bf b}_n}
\end{equation}
and $\widetilde{\bf b}_n$ is the $n$th column of $\widetilde {\bf B}$, as given in Table~\ref{tab:scheme}. With all the considered schemes, we impose ${1}/{L_s}\int_{V_s} \|{\bf j}(\vect{s})\|^2 d {\bf s} = P_s$
which yields $P = (\kappa Z_0)^2 P_s$. Notice that $P_s$ is measured in A$^2$ and $P$ in [V$^2$/m$^2$]. 
We assume $P_s = 10^{-7}$\,[A$^2$] and $P/\sigma^2_{{\rm emi}} = 90$\,dB.

\begin{figure}[t!]\vspace{-0.4cm}
\begin{center}
\includegraphics[width=0.43\textwidth]{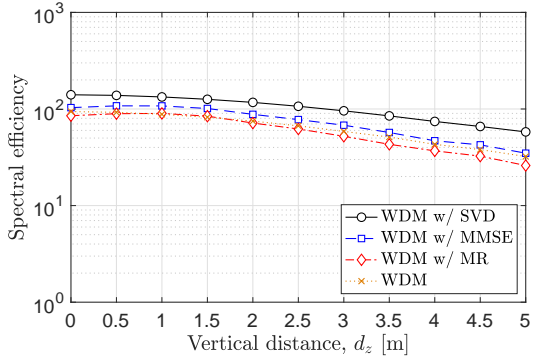}
\caption{SE in bits per channel use vs. $d_z$, with $d_x=5$\,m, and $\vartheta_s=\varphi_s=0^{\circ}$.}
\label{FigNR1}\vspace{-0.6cm}
\end{center}
\end{figure}

We begin by evaluating the impact of the vertical misalignment in Fig.~\ref{figure_linear_arrays}. Fig.~\ref{FigNR1} illustrates the spectral efficiency, in bits per channel use, {of the schemes in Table~\ref{tab:scheme}} as a function of $d_z$, for $d_x=5$\,m and $\vartheta_s=\varphi_s=0^{\circ}$. As expected, the highest SE of WDM is achieved with SVD, followed by MMSE, MR and the simplest implementation. Moreover, the SE of all the schemes decreases as $d_z$ increases, due to the lower received power as the receive segment moves up or down from the broadside alignment. This is confirmed by the fact that the SE degradation is (almost) independent of the particular scheme, meaning that it does not depend on the interference.

\begin{figure}[t!]
\begin{center}
\includegraphics[width=0.43\textwidth]{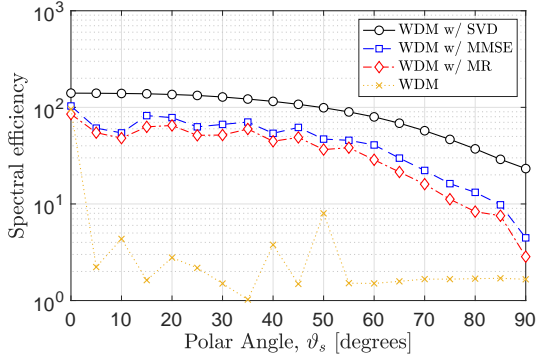}
\caption{SE in bits per channel use vs. $\vartheta_s$, for $d_x=5$\,m, $\varphi_s=0^{\circ}$ and $d_z=0$\,m.}\vspace{-0.5cm}
\label{FigNR2}
\end{center}
\end{figure}

\begin{figure}[t!]\vspace{-0.4cm}
\begin{center}
\includegraphics[width=0.43\textwidth]{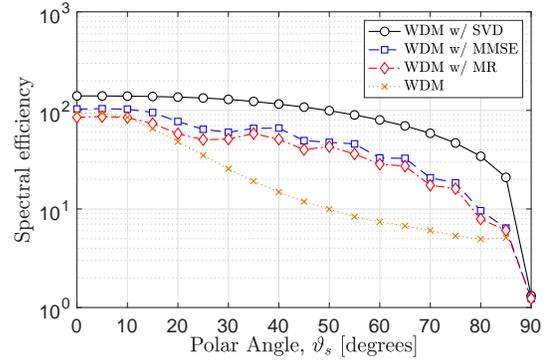}
\caption{SE in bits per channel use vs. $\vartheta_s$, for $d_x=5$\,m, $\varphi_s=90^{\circ}$ and $d_z=0$\,m.}
\label{FigNR3}\vspace{-0.6cm}
\end{center}
\end{figure}

Fig.~\ref{FigNR2} plots the SE as a function of the polar angle $\vartheta_s$, for $\varphi_s=0^{\circ}$ and $d_z=0$\,m. Unlike Fig.~\ref{FigNR1}, we see that the impact on the SE is largely different depending on the considered scheme. {Specifically, the SVD is only marginally affected for values of $\vartheta_s \le 45^\circ$ while MMSE and MR experience both some degradation that fluctuates as $\vartheta_s$ increases.} The SE with the simplest implementation of WDM is strongly compromised. Particularly, it reduces substantially as $\vartheta_s\ge 0^{\circ}$. All this can be explained by observing that varying $\vartheta_s$ increases the interference among the communication modes. While the SVD is optimal and thus performs well against the arising interference, all the other schemes are suboptimal, and 
are affected (although differently) by its presence. Similar conclusions can be drawn from Fig.~\ref{FigNR3}, where the SE is plotted in the same setting of Fig.~\ref{FigNR2} but with $\varphi_s=90^{\circ}$. {Compared to the case $\varphi_s=0^{\circ}$, the better performance of WDM can be explained with an interference reduction among the spatial modes. The other schemes behave similarly.}

\begin{figure}
\begin{center}
\includegraphics[width=0.43\textwidth]{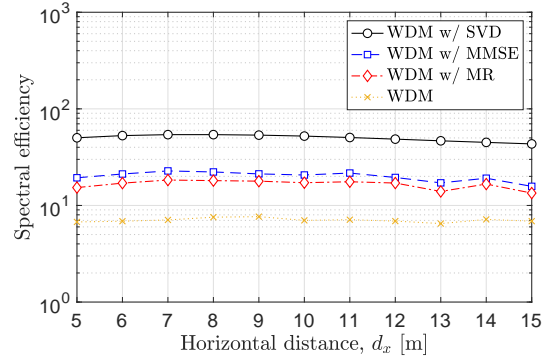}
\caption{SE in bits per channel use vs. $d_x$, for $d_z=5$\,m, $\vartheta_s$ uniformly distributed in $(0,30^{\circ})$ and $\varphi_s \in \{0,22.5^{\circ},45^{\circ},77.5^{\circ},90^{\circ}\}$.}\vspace{-0.7cm}
\label{FigNR4}
\end{center}
\end{figure}

Fig.~\ref{FigNR4} shows the \textit{average} SE as a function of the horizontal distance $d_x$ with $d_z=5$\,m. We assume that $\varphi_s$ can take any value in the set $\{0,22.5^{\circ},45^{\circ},77.5^{\circ},90^{\circ}\}$, while $\vartheta_s$ is uniformly distributed in the set $(0,30^{\circ})$. We see that the average SE is not significantly affected when $d_x$ varies from $5$\,m to $15$\,m. This can be explained by observing that the received power associated to \textit{each single mode} can increase or decrease depending on $d_x$ value. The results in Fig.~\ref{FigNR4} suggest the \textit{total} power and interference levels do not change significantly for the considered values of $d_x$.

\section{Conclusion}
This letter extended the analysis of the WDM scheme presented in~\cite{sanguinetti2022WDM} to more general scenarios in which the transmit and receive segments are arbitrarily oriented and positioned. We focused on the spectral efficiency with different digital processing architectures and showed that it is marginally affected with SVD. The simplest implementation of WDM performs also sufficiently well for practical values of geometric parameters. The analysis provided in this letter can be used to extend the application of WDM to a multiuser scenario.


\bibliographystyle{IEEEtran}
\bibliography{IEEEabrv,refs}




\end{document}